\documentclass[twocolumn,oldversion]{aa}
\usepackage{graphicx}
\usepackage{txfonts}
\input{psfig.sty}
\begin{document}

\title{Age constraints on the cosmic equation of state}


   \author{M. A. Dantas \inst{1}
          J. S. Alcaniz \inst{1} 
          \and
          Deepak Jain \inst{2}
\and
Abha Dev \inst{3}  }

   \offprints{J. S. Alcaniz}

   \institute{Observat\'orio Nacional, Rua Gal. Jos\'e Cristino 77, 20921-400 Rio de Janeiro - RJ,
Brasil\\ {\email{aldinez@on.br}}, \email{alcaniz@on.br}
         \and Deen Dayal Updhyaya College, University of Delhi , Delhi 110 015., India\\ \email{deepak@physics.du.ac.in}
\and Miranda House, University of Delhi, Delhi - 110007,
India}
      

   \date{Accepted for publication in Astronomy and Astrophysics -- January 9th}

\abstract{Dark energy is the invisible fuel that seems to drive the current acceleration of the Universe. Its presence, which is inferred from an impressive convergence of high-quality observational results along with some sucessful theoretical predictions, is also supported by the current estimates of the age of the Universe from dating of local and high-$z$ objects. In this work we study observational constraints on the dark energy equation of state ($w$) from lookback time measurements of high-$z$ galaxies, as recently published by the Gemini Deep Deep Survey (GDDS). In order to build up our lookback time sample from these observations we use 8 high-$z$ galaxies in the redshift interval $1.3 \leq z \leq 2.2$ and assume the total expanding age of the Universe to be $t_{0}^{obs} = 13.6^{+0.4}_{-0.3}$ Gyr, as obtained from current cosmic microwave background data. We show that these age measurements are compatible with values of $w$ close to $-1$, although there is still space for quintessence ($w > -1$) and phantom ($w < -1$) behaviors. In order to break possible degeneracies in the $\Omega_{\rm{m}} - w$ plane, we also discuss the bounds on this parametric space when GDDS lookback time measurements are combined with the most recent SNe Ia, CMB and LSS data.

   }

   \maketitle
%
\section{Introduction}
\label{sec:into}

Over the past decade, an increasing number of observational results have consistently indicated that we live in a nearly flat, accelerating universe composed of $\simeq 1/4$ of pressureless matter (barionic + dark) and $\simeq 3/4$ of an exotic component with large negative pressure, usually called dark energy or quintessence. The basic set of observations supporting such idea includes distance measurements to intermediary and high-$z$ Type Ia Supernovae [SNe Ia] (Riess et al. 1998; 2004; Perlmutter et al. 1999; Astier et al. 2006), measurements of the Cosmic Microwave Background (CMB) anisotropies (de Bernardis et al. 2000; Spergel et al. 2003; 2006) and the current observations of the Large-Scale Structure (LSS) in the Universe (Tegmark et al. 2004; Eisenstein et al. 2005; Cole et al. 2005). Although it is widely believed that the existence of this dark energy component provides the key and remaining pieces of information connecting the inflationary flatness prediction ($\Omega_{\rm{Total}} = 1$) with astronomical observations, nothing but the fact that it has a negative equation of state (EoS) $w \equiv p/\rho$  and that its energy density and pressure are of the order of the critical density (i.e., $p \simeq -\rho \simeq -10^{-29} \rm{g/cm^{3}}$) is known thus far (for recent reviews on this topic, see Peebles and Ratra, 2003; Padmanabhan 2004; Copeland et al. 2006). 

On the other hand, it is also well true that there is now a flurry of activity focused on unveiling the nature of dark energy or constraining any alternative explanation for the cosmic acceleration [e.g., extra dimensions effects (Sahni and Shtanov, 2003; Zhu and Alcaniz, 2005)]. However, the absence of a natural guidance from particle physics theories on the origin of this dark component seems to make clear that in order to better understand its actual nature, an important strategy is to find new observational methods or to revive old ones that could directly or indirectly quantify the amount of dark energy present in the Universe, as well as determining its EoS. In this regard, the possibility of constraining cosmological parameters from age estimates of high-$z$ objects constitutes an important and interesting attempt. In reality, since the days pre-dark energy, estimates of the age of the Universe ($t_{\rm{U}}$) have been one of the most pressing piece of data
supporting such an idea (Krauss and Turner, 1995), in that dark energy helps explain the current dating of globular clusters by allowing a period of cosmic acceleration and leading to a larger expansion age\footnote{For the widely accepted current value of the Hubble parameter, i.e., $H_o \simeq  72$ $\rm{km.s^{-1}.Mpc^{-1}}$ (Freedman et al. 1998), no flat CDM model without dark energy (whose age prediction is $H_ot_{\rm{U}} = 2/3$) may be compatible either with the direct age estimates from globular clusters or with the indirect age estimates, as provided by SNe Ia and CMB measurements (Tonry 2002; Spergel et al. 2003; 2006).}. In this regard, it also worth emphasizing that the evolution of the age of the Universe with redshift ($dt_{\rm{U}}/dz$) differs from scenario to scenario, which means that models that are able to explain  the total expanding age (at $z = 0$) may not be compatible with age estimates of high-$z$ objects (and vice-versa). This in turn reinforces the idea that dating of objects at high-redshift constitutes one of the most powerful methods for constraining the age of the Universe at different stages of its evolution (Dunlop et al. 1996; Kennicutt Jr. 1996; Spinrad et al. 1997; Hasinger et al. 2002), the first epoch of galaxy/quasar formation (Alcaniz \& Lima 2001; Alcaniz et al. 2003), as well as for discriminating among different dark energy models (Krauss 1997; Alcaniz \& Lima 1999; Jimenez \& Loeb 2002; Jimenez et al. 2003; Fria\c{c}a et al. 2005; Jain and Dev, 2006).

Recently, with the advent of large telescopes and new technics, it was possible to estimate more precisely ages of high-$z$ objects, including galaxies, quasars and galaxy clusters. In this regard, of particular interest to test current dark energy parametrizations are the discoveries of the prototypical evolved red galaxies LBDS 59W091 and LBDS 59W069 (Dunlop et al. 1996; Spinrad et al. 1997; Bruzual and Magris 1997; Dunlop 1999; Yi et al. 2000; Nolan et al. 2001), the quasar APM 08279+5255 (Hasinger et al. 2002; Fria\c{c}a et al. 2005) and the recently released sample of old passive galaxies from Gemini Deep Deep Survey [GDDS] (Abraham et al., 2004; McCarthy et al., 2004). In particular, the GDDS sample consists of 20 objects lying in the interval $1.3 \leq z \leq 2.2$, whose integrated light is dominated by evolved stars. As discussed by McCarthy et al. (2004), the data seems to indicate that the most likely star formation history is that of a single burst of duration less than 0.1 Gyr, although in some cases the duration of the burst is consistent with 0 Gyr, which means that the galaxies have been evolving passively since their initial burst of star formation. From the theoretical viewpoint, these data, along with other age estimates of high-$z$ objects, were recently used to reconstruct the shape and redshift evolution of the dark energy potential (Simon et al., 2005), as well as to place bounds on holography-inspired dark energy scenarios (Yi and Zhang, 2006).

In the present paper, by following the methodology presented by Capozziello et al. (2004) [see also Saini et al. 2000 and Pires et al. 2006], we discuss quantitatively how GDDS age estimates of high-$z$ galaxies constrain the EoS describing the dark energy. In order to perform our analysis, we transform the selected GDDS observations into lookback time (LT) measurements by assuming the total expanding age of the Universe to be $t^{obs}_o = 13.6^{+0.4}_{-0.3}$ Gyr, as recently obtained from an analysis involving the most recent CMB data (MacTavish et al., 2005). To better constrain the  dark energy EoS and energy density, we also investigate the bounds on these quantities when GDDS LT measurements are combined with the most recent SNe Ia and LSS data.

\section{Background Equations and Lookback Time}

With the assumption that the effective equation of state, $w \sim \int{w(z)\Omega_x(z)dz/\Omega_x(z)dz}$, is a good approximation for a wide class of dark energy scenarios, we start our analysis by considering a homogeneous, isotropic, spatially flat universe described by the Friedmann-Robertson-Walker line element, $ds^2 = dt^2 - a^2(t)(dx^2+dy^2+dz^2),$ where $a(t)$ is the cosmological scalar factor and we have set the speed of light $c = 1$. 

The Friedmann equation for such a scenario is given by
\begin{eqnarray}
{\cal{H}}(\mathbf{p})  =  \sqrt{\rm{E}(\mathbf{p})}\quad \mbox{with} \quad\rm{E}(\mathbf{p}) = {\Omega_m a^{-3}} + {\Omega_x a^{-3(1+w)}}
\end{eqnarray}
where ${\cal{H}}(\mathbf{p}) \equiv {\rm{H}(\mathbf{p})}/{\rm{H_0}}$, the complete
set of parameters is $\mathbf{p} \equiv \{\Omega_{j}, w\}$ ($j \equiv m$ and $x$ stand for matter and dark energy density parameters, respectively), and from now on the subscript 0 denotes present-day quantities. In this background, the lookback time-redshift relation, defined as the difference between the present age of the Universe ($t_o$)  and its age
($t_z$) when a particular light ray at redshift $z$ was emitted, can be written as
\begin{equation} \label{looktheo}
t_L(z;\mathbf{p}) = {\rm{H}}^{-1}_o \int_o^z{\frac{dz'}{(1 + z'){{\cal{H}}(\mathbf{p})}}},
\end{equation}
where ${\rm{H}}^{-1}_0 = 9.78h^{-1}$ Gyr and $h$ ranges in the interval $0.64 \leq h \leq 0.8$, as provided by the HST key project (Friedman et al., 2002). 

To proceed further, let us now consider an object (e.g., a galaxy, a quasar or a galaxy
cluster) at redshift $z_i$ whose the age $t(z_i)$ is defined as the difference between
the age of the Universe at $z_i$ and the one when the object was born (at its formation
redshift $z_F$),  i.e.,
\begin{eqnarray}
t(z_i)  =  {1 \over {\rm{H_0}}} \left[\int_{z_i}^{\infty}{\frac{dz'}{(1 +
z'){{\cal{H}}(\mathbf{p})}}} -  \int_{z_F}^{\infty}{\frac{dz'}{(1 +
z'){{\cal{H}}(\mathbf{p})}}}\right]
\end{eqnarray}
or, equivalently,
\begin{equation}
t(z_i)  = t_L(z_F) - t_L(z_i).
\end{equation}
From the above expressions, we can define the observed lookback time to an object at $z_i$ as
\begin{eqnarray} \label{lookobs}
 t^{obs}_L(z_i; \tau) & = & t_L(z_F) - t(z_i)  \nonumber \\ & &
= [t^{obs}_o - t(z_i)] - [t^{obs}_o - t_L(z_F)] \nonumber \\ & &
= t^{obs}_o - t(z_i) - \tau,    
\end{eqnarray}
where $\tau$ stands for the {incubation time} or {delay factor}, which accounts
for our ignorance about  the amount of time since the beginning of the structure
formation in the Universe until the formation time ($t_f$) of the object [see also Capozziello et al. (2005) and Pires et al. (2006) for a similar discussion]. 

\begin{figure*}
\label{fig:trans}
\centerline{\psfig{figure=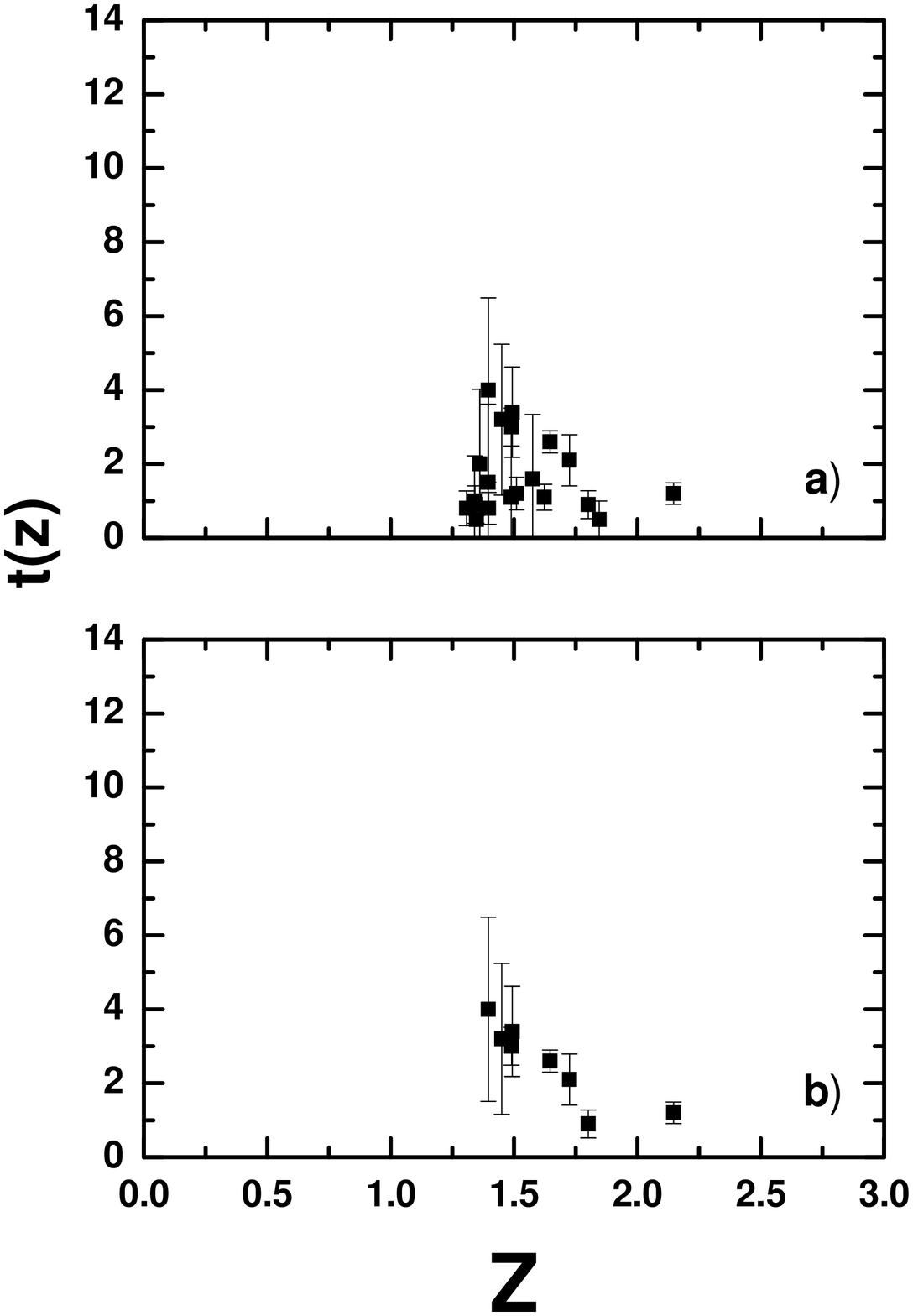,width=3.4truein,height=3.4truein,angle=0}
\psfig{figure=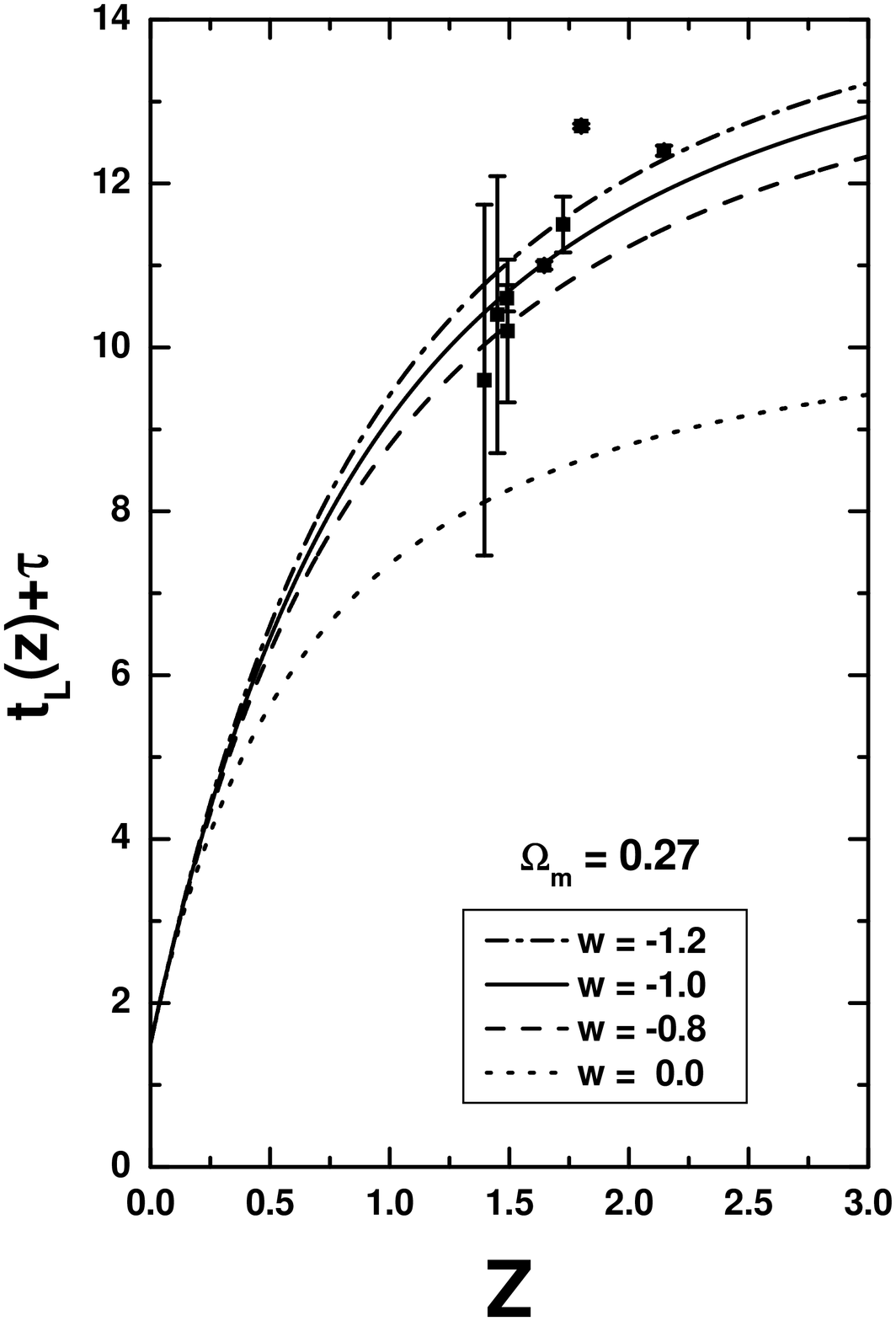,width=3.4truein,height=3.4truein,angle=0}
\hskip 0.1in} 
\caption{{\bf Left:} The age-redshift data points. {\it (a)} Original data from GDDS. This sample corresponds to 20 old passive galaxies distributed over the redshift interval $1.308 \leq z \leq 2.147$, as given by McCarthy et al. (2004). {\it (b)} The 8 high-$z$ measurements selected after the criterion discussed in the text. {\bf Right:} Lookback time relation as a function of the redshift  for some values of the EoS parameter $w$ and a fixed value of $\Omega_{\rm{m}} = 0.27$. To plot these curves, we have used  $\tau = 1.5$ Gyr. Thick line corresponds to the $\Lambda$CDM case ($w = -1$).}
\end{figure*}

\section{Lookback time Data and Statisical analysis}

\subsection{Data}

In this section we discuss quantitatively how the GDDS LT estimates of high-$z$ galaxies constrain the EoS describing the dark energy through a statistical analysis of the data. The total GDDS sample presented by McCarthy et al. (2004) consists of 20 old passive galaxies distributed over the redshift interval $1.308 \leq z \leq 2.147$ and reveal unambiguous evidence for old and metal-rich galaxies over the full redshift range. To determine the galaxy ages McCarthy et al. (2004) compare the observed spectral energy distribution with a set of synthetic spectra computed with P\'EGASE.2 (Fioc \& Rocca-Volmerange, 1997) and construct a multidimensional $\chi^2$ surface spanning a wide range of star formation histories, extinction and metalicities. Although this galaxy sample has been independently reanalyzed by Simon et al. (2005), who obtained ages within 0.1 Gyr of the GDDS collaboration estimates, it is important to emphasize that these age estimates are not free of observational and theoretical uncertainties (for instance, those related to the population synthesis models, the amount of active evolution, among others.) [As an example, see the debate involving the age estimates for the radio galaxy LBDS 59W069; Bruzual and Magris, 1997; Yi et al, 2000; Nolan, Dunlop and Jimenez, 2001].

In order to build up our LT sample, we first select from GDDS observations what we will consider as the most appropriate data to our cosmological analysis. Since our primary objective here is to impose limits on the behavior of the cosmic EoS as restrictive as possible from these high-$z$ age measurements, we adopt the criterion that given two objects at (approximately) the same $z$, the oldest one is always selected. This is justified by the fact that a universe model that is capable of explaining the existence of a very old object at a given $z$  can also naturally explain the age of the youngest objects at that $z$. By following this criterion we end up with a sample of 8 data points, which are shown in Fig. 1(b) along with the original data from GDDS collaboration [Fig. 1(a)]. The other important aspect to build up our lookback time sample [see Eq. (5) above] concerns the total age of the Universe, which is assumed in our analysis to be  $t^{obs}_o = 13.6^{+0.4}_{-0.3}$ Gyr, as obtained by MacTavish {\it et al.} (2005) from a joint analysis involving only data of the most recent CMB experiments (WMAP, DASI, VSA, ACBAR, MAXIMA, CBI and BOOMERANG) \footnote{Note that, to avoid double counting of information with SNe Ia and LSS analyses performed in the next sections, we have adopted age estimates obtained from combinations of CMB experiments only.}.

\subsection{Statistical Analysis}

In order to estimate the best-fit to the set of parameters $\mathbf{p}$ we define the likelihood function
\begin{equation}
{\cal{L}}_{age} \propto \exp\left[-\chi_{age}^{2}(z;\mathbf{p},\tau)/2\right],
\end{equation}
where $\chi_{age}^{2}$ is given by
\begin{eqnarray} \label{chi2} 
\chi_{age}^{2} &  = & \sum_{i}{\frac{\left[t_L(z_i;\mathbf{p}) -
t^{obs}_L(z_i; \tau)\right]^{2}} {\sigma_{\rm{T}}^{2}}} + \nonumber \\ & &
\quad \quad \quad \quad   + \frac{\left[t_o(\mathbf{p}) -
t^{obs}_o\right]^{2}}{\sigma_{t^{obs}_o}^{2}}.
\end{eqnarray}
Here, $\sigma_{\rm{T}}^{2} \equiv \sigma_i^{2} + \sigma_{t^{obs}_o}^{2}$, $\sigma_i^{2}$ is the uncertainty in the individual lookback time to the i$^{\rm{th}}$ galaxy of our sample and $\sigma_{t^{obs}_o} = 0.35$ Gyr stands for the uncertainty on the total expanding age of the Universe ($t^{obs}_o$). As mentioned earlier, the evolution of the age of the Universe with redshift may differ considerably from scenario to scenario, which amounts to saying that it is possible that cosmological models that are able to explain age estimates of high-$z$ objects may not be compatible with the total expanding age at $z = 0$ (Fria\c{c}a et al. 2005). In our statistical analysis, this is taken into account by including a prior on the total age of the Universe, which is equal (as should be) to the value adopted to build up the lookback time sample, i.e, $t^{obs}_o = 13.6^{+0.4}_{-0.3}$ Gyr. This in turn amounts to saying that our LT analysis is always performed with the two terms of Eq. (\ref{chi2}). 

Another important aspect concerns the delay factor $\tau$. Note that in principle there must be variations in the value of $\tau$ for each object in the sample (galaxies form at different epochs). Here, however, since we do not know the formation redshift for each object, the delay factor  is assumed as a ``nuisance" parameter, so that we marginalize over it. Note also that this marginalization may also be analytically obtained by defining a  modified log-likelihood function $\widetilde{\chi}^{\,2}$, i.e., 
\begin{eqnarray}
\widetilde{\chi}^{\,2} & = & -2\ln\int_{0}^{\infty}d{\tau}\exp
\left(-\frac{1}{2}\chi_{age}^{2}\right)\\
& = & A -\frac{{B}^2}{C}+D-2\ln\left[\sqrt{\frac{\pi}{2C}}{\rm{erfc}}\left(\frac{B}{\sqrt{2C}}\right)\right],\nonumber
\end{eqnarray}
where
\begin{eqnarray}
A=\sum_{i=1}^n\frac{\Delta^2}{\sigma_{\rm{T}}^{2}}, \quad \quad  B=\sum_{i=1}^n\frac{\Delta}{\sigma_{\rm{T}}^{2}}, \quad \quad    C=\sum_{i=1}^n\frac{1}{\sigma_{\rm{T}}^{2}}, \nonumber
\end{eqnarray}
$D$ is the second term of the rhs of Eq. (\ref{chi2}),
\begin{eqnarray}
\Delta = t_L(z_i;\mathbf{p}) - [t^{obs}_o - t(z_i)], \nonumber
\end{eqnarray}
and ${\rm{erfc}}(x)$ is the complementary error function of the variable $x$.

\begin{figure}
\label{fig:trans}
\centerline{\psfig{figure=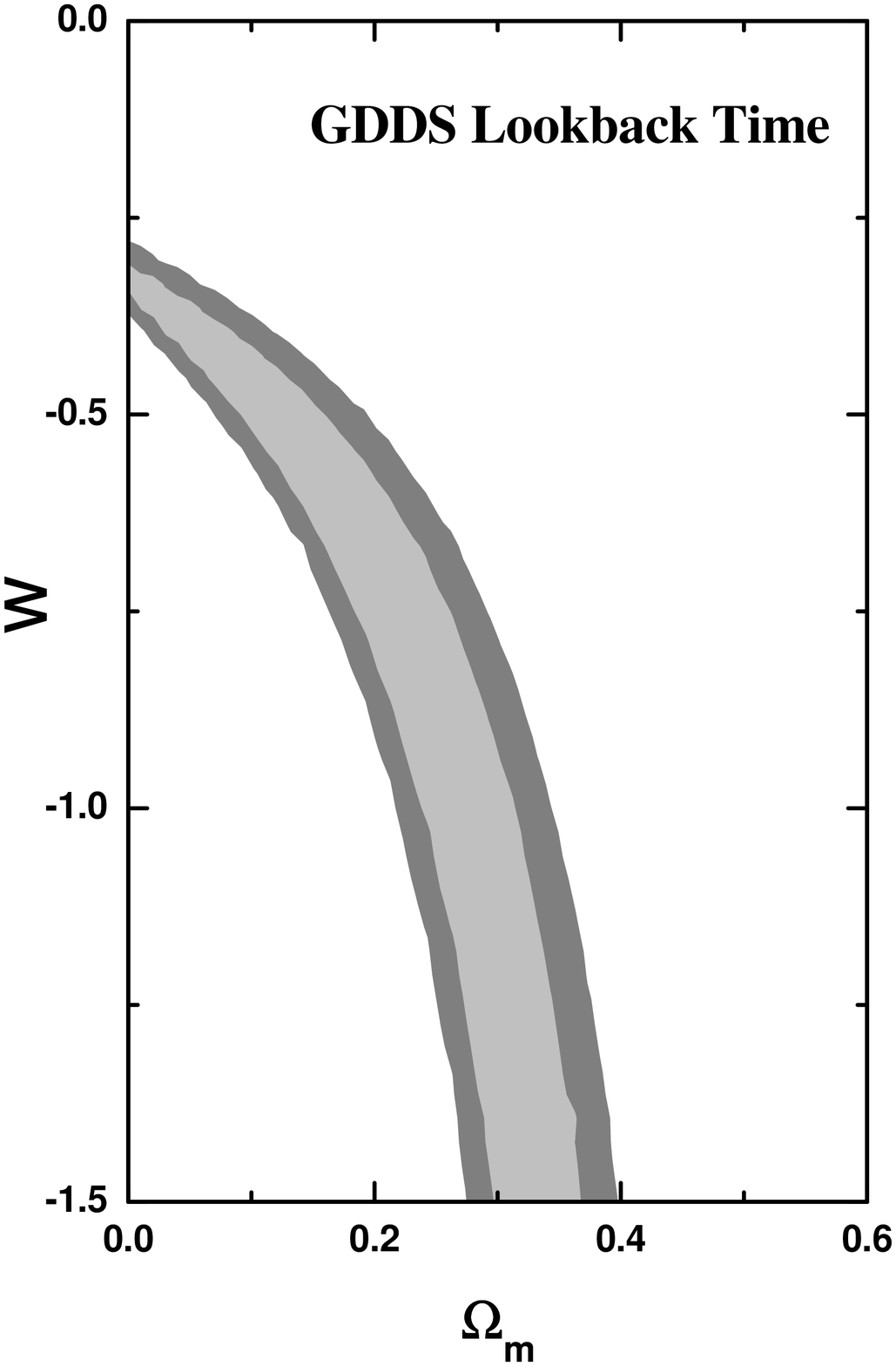,width=3.4truein,height=3.1truein,angle=0}
\hskip 0.1in} 
\caption{First results of our statistical analyses. GDDS LT constraints on the $\Omega_{\rm{m}}-w$ plane. The contours correspond to 68.3\% and 95.4\% confidence levels. The best-fit model for this analysis are  $\Omega_{\rm{m}} = 0.21$ and $w = -0.72$.}
\end{figure}

\section{Results}

Figure 1(c) shows the GDDS LT data as a function of the redshift for some selected values of the EoS parameter and a fixed value of $\Omega_{\rm{m}} = 0.27$. To plot these curves we have assumed a delay factor of $\tau = 1.5$ Gyr. Note that, in agreement with other independent analyses, only for extreme (and unrealistic) values of the delay factor (e.g., $\tau > 4.5$ Gyr), a matter-dominated universe ($w = 0$) would be still compatible with the GDDS LT estimates. 

In Fig. 2 we show the first results of our statistical analysis. Confidence regions (68.3\% and 95.4\% c.l.) are shown in the parametric space $\Omega_{\rm{m}}-w$ from the GDDS LT data discussed above. Similarly to the present constraints from SNe Ia (see, e.g., Astier et al., 2006), we note that a large interval for the dark energy EoS is allowed, whereas at 2$\sigma$ level the matter density parameter is more restricted here ($\Omega_{\rm{m}} \leq 0.35$) than in the current SNe Ia analyses ($\Omega_{\rm{m}} \leq 0.5$). Note  also that, although compatible with a supernegative behavior of the dark energy EoS ($w < -1$), the best-fit value for $w$ is much less negative ($w = -0.72$) than the one predicted by the so-called phantom scenarios (see, e.g., Caldwell, 2002; Faraoni, 2002; Alcaniz, 2004  Feng, Wang and Zhang, 2005 and references therein). For the sake of comparison, we add that If we had used the total GDDS sample of 20 galaxies (instead of applying and using the criteria discussed earlier to reduce the number of objects), our best-fit result would be $\Omega_{\rm{m}} \simeq 0$ and $\omega = -0.32$, which is very far from any realistic estimate using the current sets of observational data. It is also worth observing that GDDS LT data provide contours in the $\Omega_{\rm{m}}-w$ plane very  similar to those arising from distance meausurements from SNe Ia observations (see Fig. 3b), which means that a joint analysis involving these data sets may also be understood as an extended one, in the sense that while SNe Ia data reach $z \leq 1.01$ (Astier et al., 2006), GDDS data lies in the redshift interval $1.308 \leq z \leq 2.147$.
 

\begin{figure*}
\label{fig:trans3}
\centerline{\psfig{figure=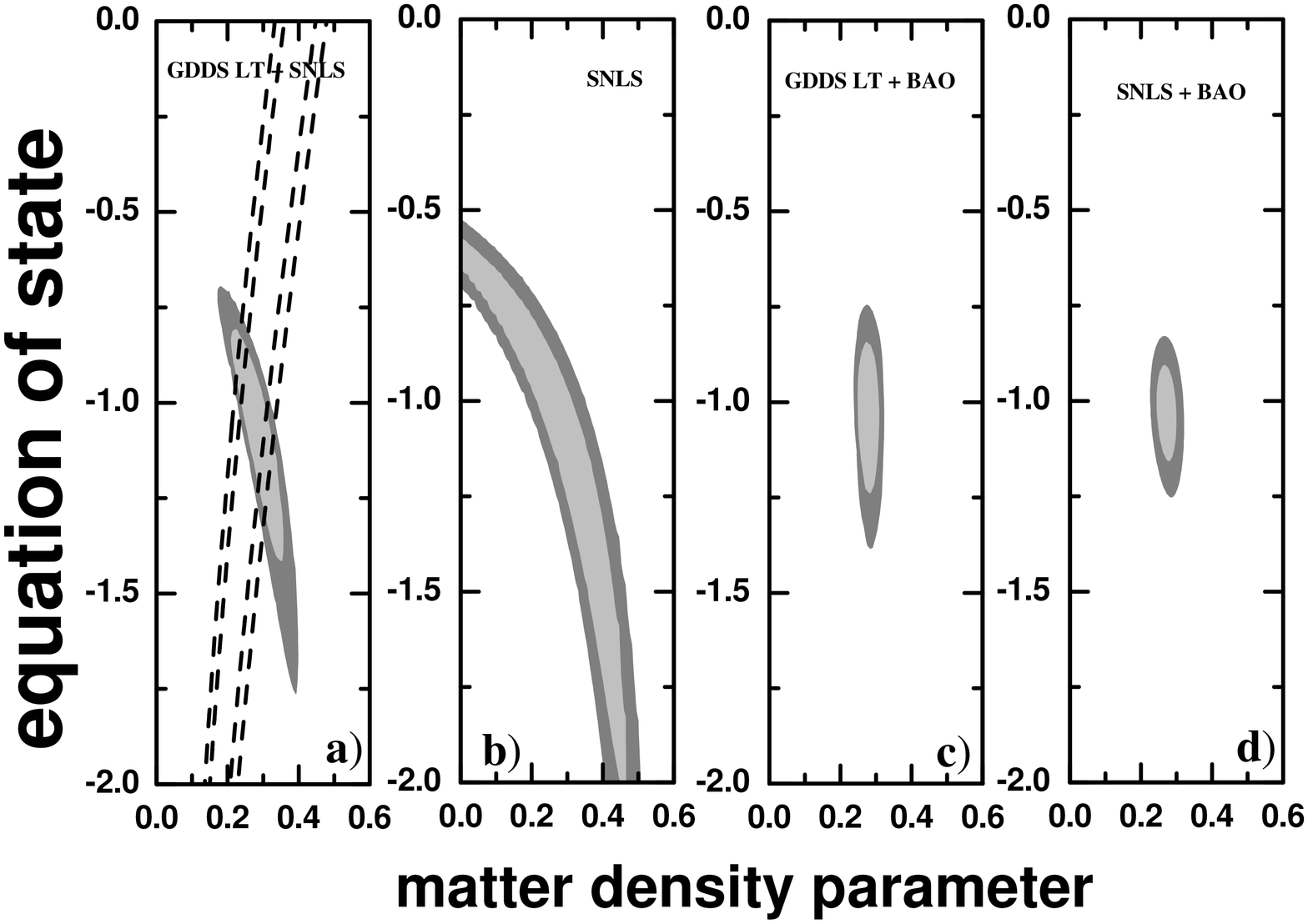,width=7.4truein,height=4.4truein,angle=-90}
\hskip 0.1in}
\caption{Complementary results of our statistical analyses. {\bf a)} GDDS LT plus SNLS constraints on the $\Omega_{\rm{m}}-w$ plane. The contours correspond to 68.3\% and 95.4\% confidence levels. The best-fit model for this analysis are  $\Omega_{\rm{m}} = 0.28$ and $w = -1.05$ with $\chi^2/\nu = 1.02$ . The dotted lines represent the constraints from BAO measurements. {\bf b)} SNLS constraints on the parametric space $\Omega_{\rm{m}}-w$ shown here for the sake of comparison (see also Astier et al. 2006). {\bf c)} The same as in Panel (a) when GDDS LT data are combined with BAO measurements. {\bf d)} Current constraints on the $\Omega_{\rm{m}}-w$ plane from SNLS + BAO data.}
\end{figure*}

\subsection{Complementary Analysis}

In order to complement our lookback time study on the parametric plane $\Omega_{\rm{m}}-w$, we combine our GDDS LT data with some of the most recent sets of cosmological observations. We use here the current SNe Ia data from Supernova Legacy Survey (SNLS) collaboration corresponding to the first year results of its planned five years survey (Astier et al. 2006). The SNLS sample includes 71 high-$z$ SNe Ia in the redshift range $0.2 \leq z \leq 1$ and 44 low-$z$ SNe Ia compiled from the literature but analyzed in the same manner as the high-$z$ sample. This data-set is arguably the best high-z SNe Ia compilation to date and seems to be in good agreement with WMAP results in that it favors values of $w \simeq -1$ (Jassal et al. 2006).  We also include in our joint analysis current  LSS data to help break the degeneracy between the dark energy EoS and the matter density parameter.  For the LSS data, we use the recent measurements of the BAO peak in the large scale correlation function detected by Eisenstein et al. (2006) using a large sample of luminous red galaxies from the SDSS Main Sample. The SDSS BAO measurement provides ${\cal{A}} = 0.469(n_S/0.98)^{-0.35} \pm 0.017$, with ${\cal{A}}$ defined as
\begin{equation}
{\cal{A}} \equiv \frac{\Omega_{\rm{m}}^{1/2}}{z_{\rm{BAO}}}\left[z_{\rm{BAO}} \frac{\Gamma^{2}(z_{\rm{BAO}};\mathbf{p})}{{\rm{E}}(z_{\rm{BAO}};\mathbf{p})}\right]^{1/3},
\end{equation}
where $\Gamma$ is the distance coordinate, $z_{\rm{BAO}} = 0.35$, ${\rm{E}}(z_{\rm{BAO}};\mathbf{p})$ is given by Eq. (1), and we take the scalar spectral index $n_S = 0.95$, as given by Spergel et al. (2006).

Figure 3(a) shows the resulting likelihood from the GDDS LT + SNLS data. There, confidence contours at 68.3\% and 95.4\% are shown in the plane $\Omega_{\rm{m}}-w$. Note that the allowed parametric space is now considerably reduced relative to the previous panel (Fig. 2) and to Fig. 3(b) which shows, for the sake of comparison, the $\Omega_{\rm{m}}-w$ constraints from SNe Ia alone (see also Astier et al. 2006). In particular, less negative ($w \ge -0.7$) and very negative ($w \leq -1.75$) values of $w$ are excluded at 2$\sigma$ level. Note also that while the SNLS data alone cannot place any restrictive bounds on the dark energy EoS at 1 or 2$\sigma$ levels, the combination GDDS LT + SNLS data moves up the confidence regions to values of $w \simeq -1$, which is in agreement with other current sets of data (see, e.g, Jassal et al. 2006 for a discussion).

The dotted lines in Fig. 3(a) represent, respectively, the constraints from BAO measurements on the plane $\Omega_{\rm{m}}-w$. As is well known, since this quantity has been measured at a specific redshift ($z_{\rm{BAO}} = 0.35$), it forms bands on the parametric space $\Omega_{\rm{m}}-w$ instead of ellipsoids, as in the case of GDSS lookback time and SNe Ia data. Here, however, the most important aspect associated with these measurements lies in the fact that the bands of allowed parameters are roughly perpendicular to the major axis of the ellipsoid arising from the GDDS LT + SNLS data, which suggests that a combined analysis may go even further in breaking the degeneracies inherent to the parametric plane $\Omega_{\rm{m}}-w$. The results of these complementary analyses are shown in Figs. 3(c), 3(d) and 4. In the former, we show the GDDS LT + BAO constraints on the plane $\Omega_{\rm{m}}-w$. Note that a considerable part of the parametric space is now ruled out at 2$\sigma$ level [mainly relative to Panel 3(a)] and an upper (lower) limit $w < -0.75$ ($w > -1.4$) can be roughly placed on the EoS parameter. These results should be compared with the very restrictive constraints from SNLS + BAO data [Panel 3(d)], providing $\Omega_{\rm{m}} = 0.271$ and $w \simeq -1.023$ (Astier et al. 2006). As expected, when all the cosmological observables discussed in this paper are combined in a joint analysis (Fig. 4) the allowed parametric space is even more reduced relative to the previous analyses (Figs. 2, 3a-3d), with the value of dark energy EoS converging to $\simeq -1$ (although there is still space for quintessence and phantom behaviours). The best-fit parameters for this total joint analysis are $\Omega_{\rm{m}} = 0.27$ and $w = -1.02$ with reduced $\chi^2_{\nu} = 1.02$ ($\equiv \chi^2_{min}/\nu$, where $\nu$ is the numeber of degrees of freedom). This  roughly corresponds to the current standard $\Lambda$CDM scenario, i.e., an accelerating universe with deceleration parameter $q_0 = -0.57$, transition redshift (at which the expansion switches from deceleration to acceleration) $z_{\rm{T}} = 0.75$, and total expanding age of 9.6$h^{-1}$ Gyr ($\simeq 13.5$ Gyr for $h = 0.71$). At 95.4\% c.l. we also found $\Omega_{\rm{m}} = 0.27 \pm 0.02$ and $w = -1.02^{+0.12}_{-0.11}$. The best-fit parameters for all the analyses performed in this paper along with the corresponding $\chi^2_{\nu}$ are summarized in Table I.

\begin{figure}[t]
\label{fig:trans}
\centerline{\psfig{figure=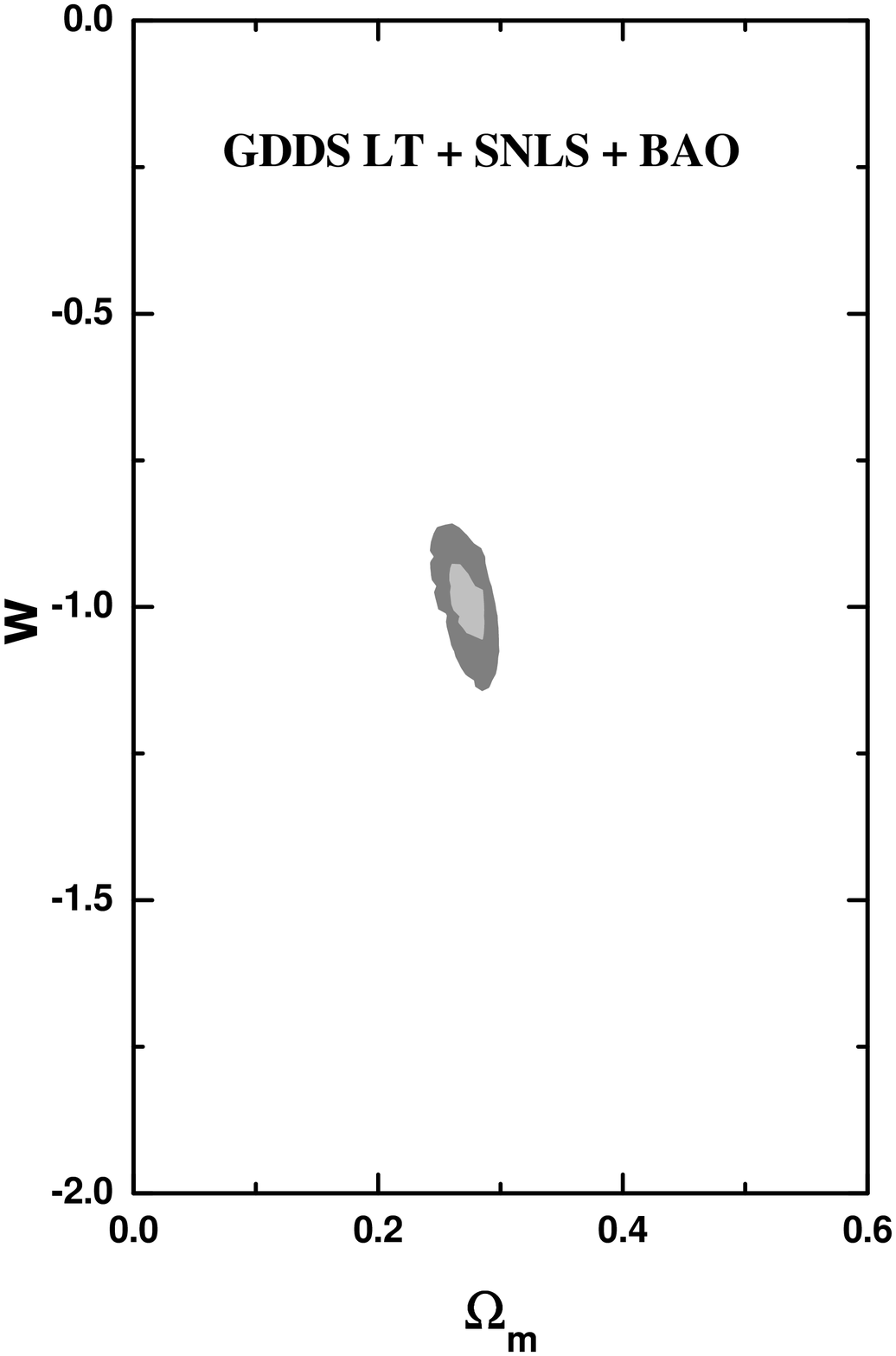,width=3.4truein,height=3.1truein,angle=0}
\hskip 0.1in} 
\caption{Joint analysis involving GDDS LT + SNe Ia + BAO data. Note that, relative to the previous Panels, the allowed parametric space is now considerably reduced. The best-fit values for all the analyses performed in this paper are presented in Table I.}
\end{figure}

\begin{table}
\caption{Best-fit values for $\Omega_{\rm{m}}$ and $w$}
\begin{tabular}{lclr}
Test& $\Omega_{\rm{m}}$  & \quad  $w$& $\chi^2/\nu$\\ 
\hline \hline \\ 
GDDS Lookback Time & 0.21  & -0.72 & 1.55 \\ 
GDDS Lookback Time + SNLS & 0.28 & -1.05 & 1.02\\ 
GDDS Lookback Time + BAO & 0.29 & -1.03 & 1.58\\
Lookback Time + SNLS + BAO & 0.27 & -1.02 & 1.02\\
\hline \hline \\
\end{tabular}
\end{table}

\section{Summary}

The recent accumulation of independent observational results has opened up a robust window for probing the behavior of the dark energy. However, as is well known, most of the methods employed to place limits on the dark energy EoS ($w$) or, more generically, on the parametric space $\Omega_{\rm{m}} - w$, are essentially based on distance measurements to a particular class of objects or physical rulers (SNe Ia, CMB, galaxy clusters, etc.). In this regard, it is also particularly important to obtain accurate and independent bounds on the physical behavior of the dark energy, as well as on the other main cosmological parameters, from physics relying on different kinds of observations.

In this paper we have followed this direction and studied constraints on the parametric space $\Omega_{\rm{m}} - w$ from age measurements of high-$z$ galaxies, as recently provided by the Gemini Deep Deep Survey. By transforming the GDSS data into lookback time estimates (by using current estimates of the total age of the Universe from LSS and CMB data), we have discussed quantitatively how the current age data constrain the EoS describing the dark energy. We have shown that, although allowing a large interval for $w$ (similarly to current SNe Ia data), these LT estimates prefer values of $w > -1$ (best-fit) and constrain the matter density parameter to be $\Omega_{\rm{m}} \leq 0.35$ at 2$\sigma$ level. The most restrictive bounds on the parametric plane $\Omega_{\rm{m}} - w$ are obtained when age and distance measurements are combined in a joint statitical analysis. In this case, using GDDS LT along with the most recent SNe Ia and LSS observations, we have found $\Omega_{\rm{m}} = 0.27 \pm 0.02$ and $w = -1.0^{+0.12}_{-0.11}$ at 95.4\% (c.l.) which, although being compatible with a quintessence and phantom cases, clearly favors the cosmological constant behavior.

Finally, we emphasize the importance of more and more precise age measurements of high-$z$ objects. These observations will certainly provide a new and complemetary tool to test the reality of the current cosmic acceleration, to place restrictive bounds on the cosmic EoS characterizing the dark energy and, as a consequence, to distingush among the many alternative world models. The present work, therefore, highlights the cosmological interest in the observational search for old collapsed objects at low, intermediary and high redshifts.

\section*{Acknowledgments}
The authors are very grateful to R. Jimenez, J. L. Kohl-Moreira and N. Pires for valuable discussions. MAD is supported by CAPES. JSA is supported by CNPq under Grants No. 307860/2004-3 and 475835/2004-2 and by Funda\c{c}\~ao de Amparo \`a Pesquisa do Estado do Rio de Janeiro (FAPERJ) No. E-26/171.251/2004.

\end{document}